\documentclass[12pt]{article}
\usepackage{graphicx}
\usepackage{fullpage}

\begin{document}
\renewcommand{\thefootnote}{\fnsymbol{footnote}}

\begin{center}
{\Large {\bf Neutralino Dark Matter in 2005\footnote{Plenary talk at
  PASCOS05, Gyeongju, Republic of Korea, June 2005}}} \\

\vspace*{5mm}

{\it Manuel Drees} \\
\vspace*{3mm}
Physikalisches Institut der Universit\"at Bonn, Nussallee 12, 53115
  Bonn, Germany

\end{center}

\begin{abstract}
I summarize some recent work on supersymmetric neutralinos as candidates for
cold Dark Matter in the Universe. This includes a new scan of mSUGRA parameter
space, with special emphasis on neutralinos annihilating predominantly through
exchange of the light CP--even Higgs boson, and on bounds on sparticle
masses. Next, prospects of testing models with TeV higgsino--like Dark Matter
at colliders are discussed. Finally, I briefly comment on extensions of the
mSUGRA model, and on scenarios with non--standard cosmology.

\end{abstract}

\clearpage

\section{Introduction}

At least within the context of Einsteinian (or, indeed, Newtonian\footnote{A
  recent attempt to construct an alternative consistent theory of gravity
  \cite{md_mond} seem to me, as a particle physicist, far more baroque than
  typical particle physics models of Dark Matter, which are occasionally
  accused of being arbitrary by traditional astronomers.}) gravity, evidence
for the existence of Dark Matter (DM) is overwhelming. This goes back to Oort
and Zwicky, who showed in the 1930's that there must be Dark Matter in our
galaxy and in distant clusters of galaxies, respectively \cite{md_history}.
Later measurements of the total DM density, using ever larger systems, as well
as improved determinations of the total baryon content of the Universe from
analyses of Big Bang nucleosynthesis, implied that most of the DM must be
non--baryonic. This was confirmed recently by analyses of data on the cosmic
microwave background (CMB) anisotropies, in particular by the WMAP satellite,
as well as other data on the large scale structure of the Universe, which
allowed to determine many cosmological parameters with unprecedented precision
\cite{md_wmap}. One should be aware, however, that these determinations are
quite indirect, and hence rely on a host of assumptions about the history of
the Universe. In particular, one has to assume a rather simple form of the
primordial power spectrum, as predicted by simple models of inflation. While
these assumptions seem rather reasonable to me (see \cite{md_sarkar} for a
more contrarian view), and lead to an overall quite satisfactory (though not
perfect) description of all existing data, it is certainly possible that these
assumptions are not quite correct. In this talk I will therefore use a rather
conservative (nominal) 3$\sigma$ error band for the quantity of interest here,
\begin{equation} \label{md_1}
0.087 \leq \Omega_{\rm DM} h^2 \leq 0.138 \,.
\end{equation}
Here $\Omega_{\rm DM}$ is the DM mass density in units of the critical
density, and $h$ is the scaled Hubble constant; observations imply $h^2 \simeq
0.5$ with $\sim 10\%$ error. The combination $\Omega h^2$ is relevant, since
this quantity can most easily be computed theoretically. 

These calculations start from the assumption that the DM particles $\chi$ once
were in full thermal (i.e., both kinetic and chemical) equilibrium with SM
particles. This requires the Universe to once have been hotter than about
$m_\chi / 20$. Since the equilibrium $\chi$ number density is suppressed
exponentially by the Boltzmann factor for $T < m_\chi$, the rate $\Gamma_\chi$
of reactions that produce\footnote{Note that only SM particles with $E \geq
  m_\chi$ can produce DM particles; the density of such energetic SM particles
  is suppressed by the same Boltzmann factor.} or destroy DM particles would
eventually have become smaller than the expansion rate $H$. At this point the
DM particles are said to decouple. Under the assumption that the Universe
evolved adiabatically after $\chi$ decoupling, one finds for today's DM
density \cite{md_kt}
\begin{equation} \label{md_2}
 \Omega_{\rm DM} h^2 \propto \frac {1} { \langle v \sigma_{\rm eff} \rangle}
 \,. 
\end{equation}
Here $\langle \dots \rangle$ denotes thermal averaging, $v$ is the relative
velocity of two annihilating $\chi$ particles in their cms frame, and
$\sigma_{\rm eff}$ is the effective annihilation cross section of two $\chi$
particles into SM particles; in the absence of co--annihilation (see below)
this is nothing but the usual annihilation cross section $\sigma(\chi \chi
\rightarrow X_{SM})$, summed over all final states $X_{SM}$ that only contain
SM particles. Note that the constant of proportionality in eq.(\ref{md_2}) is
inversely proportional to the Hubble parameter at decoupling. In standard
cosmology, decoupling occurs when the Universe was radiation dominated, so
that $H \propto g_* T^2 /M_{\rm Pl}$, where $g_*$ is the number of effective
(relativistic) degrees of freedom and $M_{\rm Pl}$ is the Planck mass. In this
case eq.(\ref{md_2}) leads to a relic density in the range (\ref{md_1}) for
roughly weak--scale cross section, i.e. $\chi$ should be a weakly interacting
massive particle (WIMP).

Under the stated reasonable assumptions -- of a sufficiently hot,
adiabatically evolving Universe containing essentially only SM particles after
the DM particles decoupled -- we have succeeded in reducing the problem of
calculating the DM relic density to a problem of calculating cross
sections. This requires choosing a particle physics model for $\chi$. Here I
will assume that $\chi$ is the lightest neutralino $\tilde \chi_1^0$, which is
the only possible DM candidate in the visible sector of the Minimal
Supersymmetric Standard Model \cite{md_book}. From the particle physics point
of view, this is the by far best motivated WIMP candidate. In the next Section
I will survey the DM--allowed regions of mSUGRA parameter space, with special
focus on two quite different scenarios: a bino--like neutralino annihilating
mostly through exchange of the light CP--even Higgs boson, and a heavy
higgsino--like neutralino. Then I will briefly comment on non--minimal
scenarios, before presenting my conclusions.

\section{Neutralino Dark Matter in mSUGRA}

The mSUGRA model \cite{md_book} allows to describe the entire spectrum of
superparticles and Higgs bosons by four continuous parameters and a sign,
\begin{equation} \label{md_3}
(m_0, \, m_{1/2}, \, A_0,\, \tan\beta, \, {\rm sign}(\mu) \, )\, .
\end{equation}
Here, $m_0$ is the common (running) soft SUSY breaking scalar mass, taken at
scale $M_X \simeq 2 \cdot 10^{16}$ GeV where the SM gauge couplings meet;
$m_{1/2}$ is the common (running) gaugino mass at that scale; $A_0$ is the
common (running) trilinear scalar soft breaking parameter, also taken at scale
$M_X$; $\tan\beta$ is the ratio of vacuum expectation values (vevs) of the two
neutral Higgs fields present in the MSSM, taken at the weak scale; and $\mu$
is the supersymmetric Higgs(ino) mass parameter\footnote{The absolute value of
  $\mu$ is fixed by the requirement of proper electroweak symmetry breaking.
  The sign of $\mu$ is not scale dependent.}. The parameters in (\ref{md_3})
determine the spectrum of physical masses through a set of renormalization
group equations, as well as ``threshold'' loop corrections to translate
running into pole masses. In the results presented below these are calculated
using the Suspect program package \cite{md_suspect}. The resulting spectrum
has to pass several constraints before being accepted as a possible
description of Nature. These include collider (chiefly LEP) limits on the
production of superparticles and Higgs bosons, radiative $b \rightarrow s
\gamma$ and $b \rightarrow s \ell^+ \ell^-$ decays, and the DM constraint
(\ref{md_1}), interpreted as described in the discussion of
eq.(\ref{md_2}). Further details can be found in \cite{md_ddk1,md_ddk2}.

\subsection{Scans of parameter space}

\begin{figure}[h!]
\rotatebox{0}{\includegraphics[height=.26\textheight]{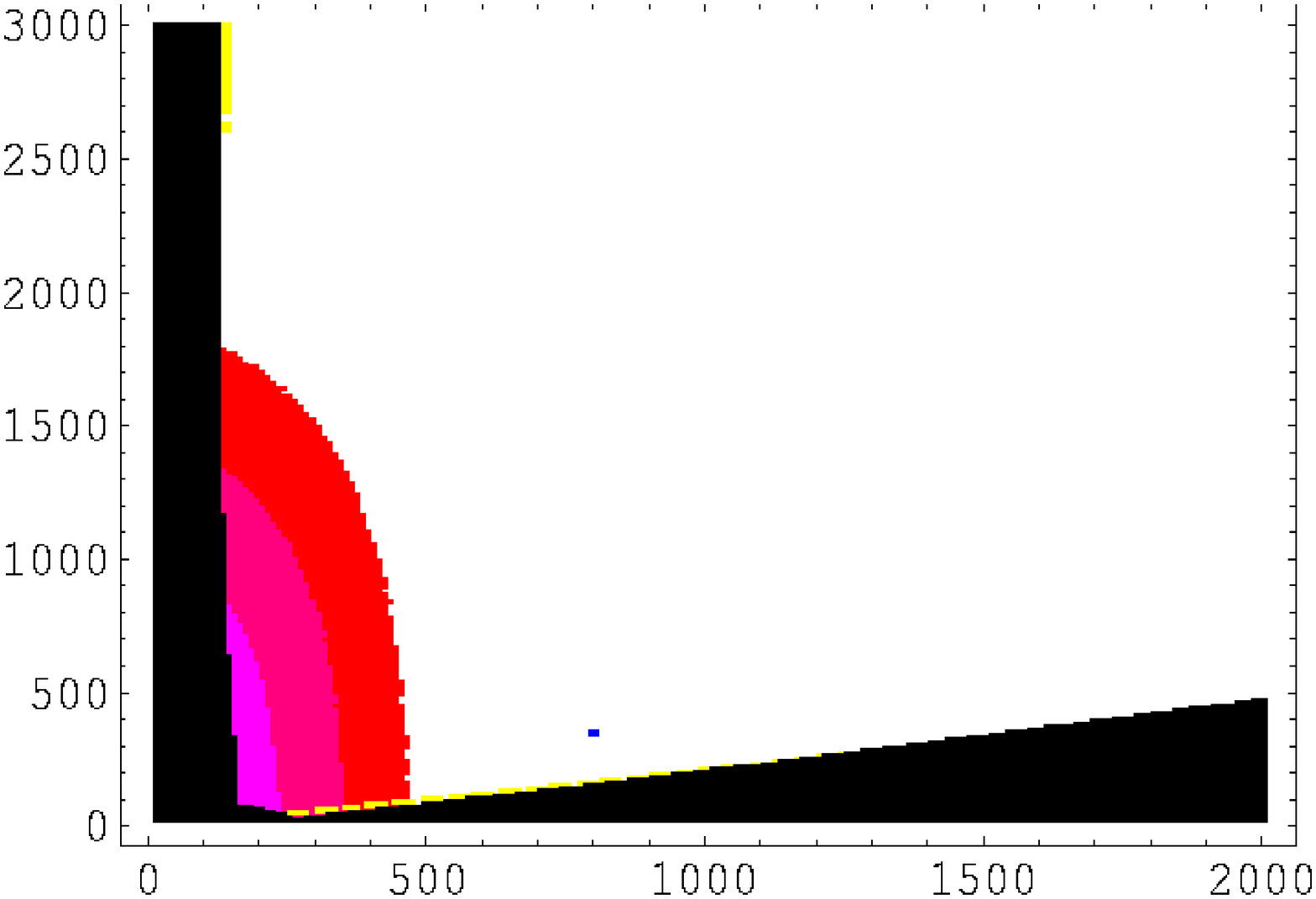}}
\rotatebox{0}{\includegraphics[height=.26\textheight]{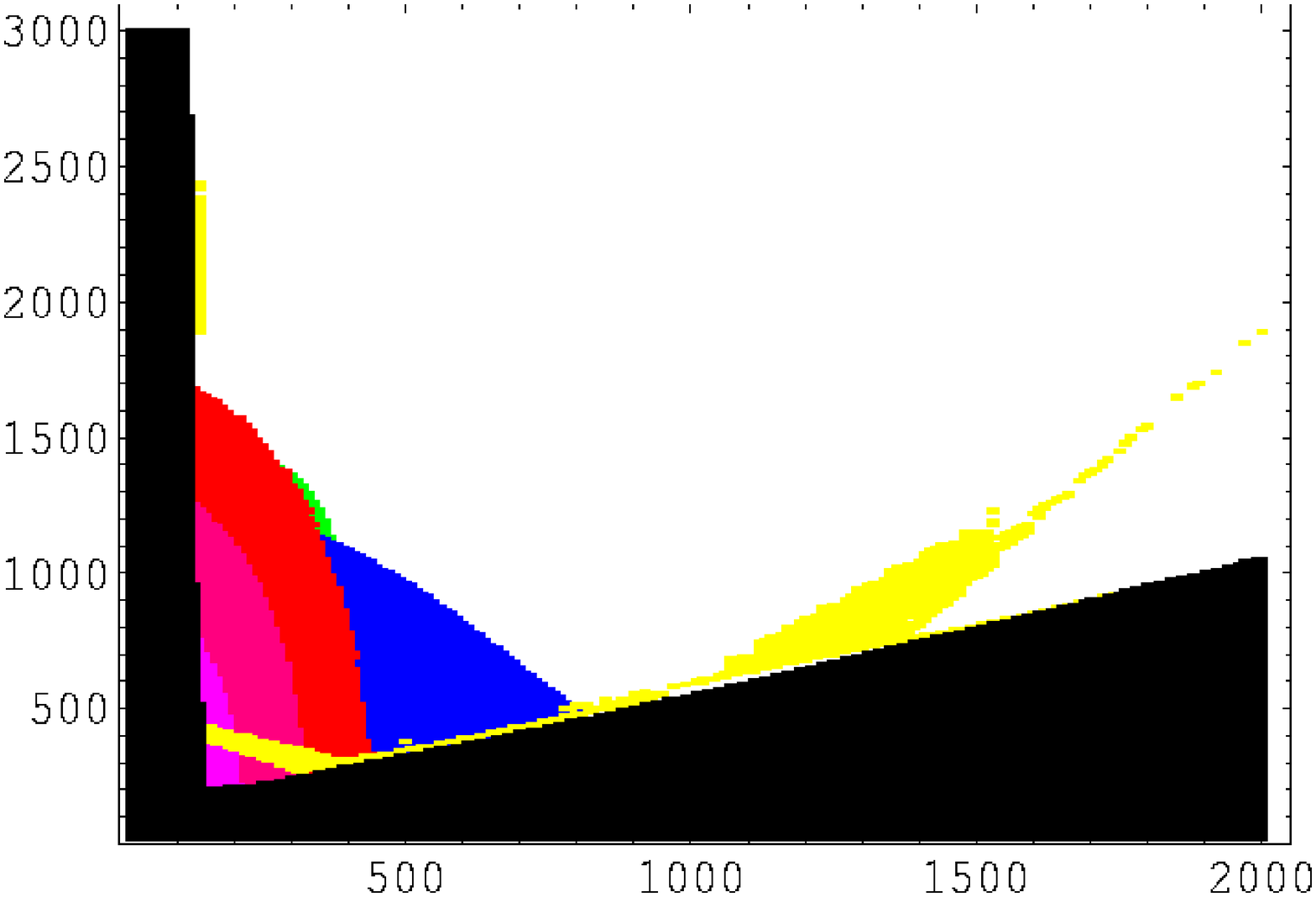}}

  \caption{The $(m_{1/2}, m_0)$ plane in mSUGRA for $\tan\beta = 10$ (left) and
    50 (right). The values of the other parameters are: $A_0 = 0, \, m_t =
    178$ GeV, and $\mu > 0$. See the text for the meaning of the different
    shaded regions.}
\end{figure}

Examples of scans of the mSUGRA parameter space are shown in Fig.~1. The black
regions are excluded by theoretical constraints (in particular, by the
requirement that the lightest superparticle, which is stable, must not be
charged), as well as by the searches for sparticles. The pink regions are
excluded by searches for neutral Higgs bosons at LEP. These LEP data show some
(weak) evidence for the existence of an SM--like Higgs boson with mass near
114 GeV; the mSUGRA regions that can explain this small excess of Higgs--like
events are shown in red. The blue region is favored by recent measurements
\cite{md_gmudata} of the anomalous magnetic moment of the muon, if data from
hadron production at $e^+e^-$ colliders are used to predict the SM
contribution; in this case a positive SUSY contribution is required at the
$\sim 2.5 \, \sigma$ level. However, if instead hadronic $\tau$ decay data are
used to evaluate the SM prediction, the data are compatible with a vanishing,
or even small negative, SUSY contribution to $g_\mu$. The green region is
excluded by the measured branching ratio for radiative $b \rightarrow s
\gamma$ decays. As argued in \cite{md_ddk2}, this bound is somewhat suspect;
it can certainly be circumvented without affecting collider (or DM)
phenomenology \cite{md_leszek}. This region is therefore shown behind, rather
than in front of, the other regions. Finally, in the yellow regions the
$\tilde \chi_1^0$ relic density lies in the desired range (\ref{md_1}). The
results shown in Fig.~1 broadly agree with those of other recent mSUGRA scans
\cite{md_other}.

One recognizes several distinct regions with correct DM relic density. In the
{\em bulk region} \cite{md_dn1} slepton masses are sufficiently small to allow
a large $\tilde \chi_1^0 \tilde \chi_1^0 \rightarrow \ell^+ \ell^-$ cross
section. However, this region is getting squeezed severely by Higgs searches
at LEP. In the {\em $\tilde \tau$ co--annihilation region} \cite{md_stau} the
$\tilde \chi_1^0 - \tilde \tau_1$ mass splitting is so small that
co--annihilation processes like $\tilde \chi_1^0 \tilde \tau_1 \rightarrow
\tau \gamma$ have to be included in the calculation of the effective LSP
annihilation cross section \cite{md_griest}, since they also reduce the total
number of superparticles in the early Universe. This increases the effective
annihilation cross section in eq.(\ref{md_2}) by about one order of magnitude.
These regions lie close to the lower black regions in Fig.~1.\footnote{A
  conceptually similar stop co--annihilation region \cite{md_stop} is
  difficult to realize in mSUGRA.} The {\em $A-$pole region} \cite{md_dn1} is
characterized by $2 m_{\tilde \chi_1^0} \simeq m_A$, so that LSP annihilation
is enhanced by the near--resonant exchange of the CP--odd Higgs boson $A$ in
the $s-$channel. In mSUGRA this can only be realized at large $\tan\beta$,
since the contribution from the $b$ Yukawa coupling, which scales like
$1/\cos\beta$, is essential for reducing $m_A$ through renormalization group
running \cite{md_dn0}. At large $m_0 \gg m_{1/2}$ one finds the {\em focus
  point} \cite{md_focus} or {\em hyperbolical branch} \cite{md_hyper} region,
where the LSP has a sizable or even dominant higgsino component. The location
of this region depends very strongly on the precise value of the top mass.

\begin{figure}
\centerline{\includegraphics[height=.4\textheight]{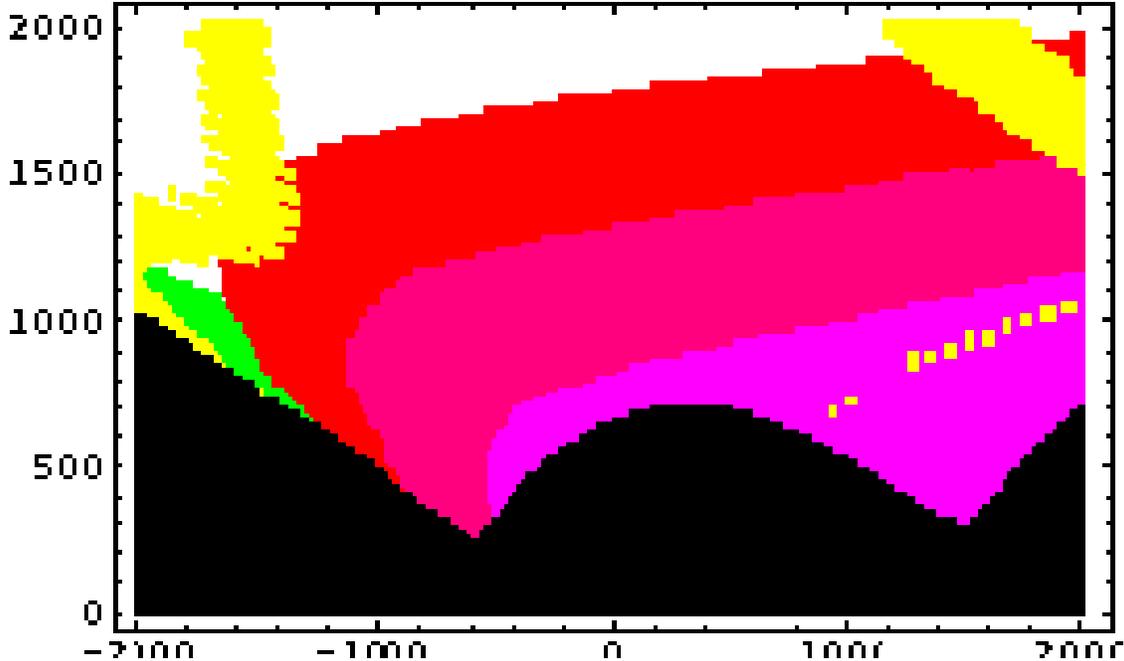}}

  \caption{The $h-$pole region in the $(A_0, m_0)$ plane of mSUGRA parameter
    space, for $m_t = 178$ GeV, $m_{1/2} = 140$ GeV, $\tan\beta = 10$ and $\mu
    > 0$. Notation is as in Fig.~1.}
\end{figure}

Finally, upon closer inspection you will see small DM--allowed regions at
$m_0 \sim 2$ to 3 TeV and $m_{1/2}$ close to its lower bound (which
essentially comes from chargino searches at LEP). In these regions of
parameter space the $\tilde \chi_1^0$ annihilation cross section is enhanced
by the near--resonant exchange of the light CP--even Higgs boson $h$. This
region has therefore been called the {\em $h-$pole region} in \cite{md_ddk2}.
It featured prominently in pre--LEP2 mSUGRA scans (see e.g. \cite{md_dn1,
  md_old}), but then was squeezed almost out of existence by Higgs and
chargino searches.  It has recently been resurrected \cite{md_baer} by
improved calculations of $m_h$, as well as by the somewhat increased central
value of the top mass (which, however, shows a very recent tendency of falling
again, if preliminary data \cite{md_top} are to be trusted). For $m_t = 178$
GeV, the $h-$ pole region requires 
\begin{equation} \label{md_4}
m_{1/2} \in [130 \, {\rm GeV}, \, 145 \,{\rm GeV}] \ \ \ \ \ \ \ (h-{\rm pole
  \ region}).
\end{equation}
However, as shown in Fig.~2, it covers quite a substantial area
in the $(A_0, m_0)$ plane for $m_{1/2}= 140$ GeV.

\subsection{Mass bounds}

Bounds on physical (pole) masses might be a more meaningful way to show the
possibilities of mSUGRA than the ubiquitous plots of allowed regions in the
space of basic input parameters (\ref{md_3}). For one thing, these plots
always fix the values of other free parameters. Usually $A_0 = 0$ is chosen,
but nonvanishing values of $A_0$ can have quite significant impact on allowed
spectra \cite{md_ddk1,md_pauss}. Similarly, the current error on the top quark
mass is still significant. Varying $m_t$ in its 90\% c.l. interval (according
to \cite{md_d0top}) moves around the lower bounds on $m_0$ or $m_{1/2}$ from
Higgs searches by hundreds of GeV; the shift of the focus point / hyperbolical
branch region is even bigger. Finally, the parameter $\tan\beta$ is still
almost completely unconstrained, apart from the lower bound $\tan\beta \geq
1.5$ that follows from Higgs searches in mSUGRA (for stop masses below 2
TeV). As well known \cite{md_dn0}, the sparticle spectrum for models with
large $\tan\beta$ looks quite different from that for low or moderate values
of $\tan\beta$, with quite drastic impact on collider physics
\cite{md_hightb}. 

One obtains the least biased view of the allowed ranges of masses by simply
scanning over the entire parameter set that is consistent with a given set of
constraints.\footnote{I do not see much value in attempts \cite{md_allanach}
  to assign probabilities to certain regions of parameter space. These
  parameters cannot be interpreted as randomly distributed quantities, unless
  one subscribes to ``landscape'' and ``multiverse'' ideas, which to my mind
  come dangerously close to pseudoscience. If these parameters are indeed
  fixed, it makes little sense to assign probability distributions to them.}
Not all constraints should be treated on an equal footing. Lower bounds on
masses (or cross sections or branching ratios) from accelerator--based
experiments are most robust, since both beam and detector are well controlled
by the experimenter (I hope). Bounds on masses and tree--level cross sections
usually also don't have many theoretical ambiguities. Bounds on processes that
only occur at quantum level, like $b \rightarrow s \gamma$ decays or the
anomalous magnetic moment of the muon, are already somewhat less robust, since
here in general several diagrams contribute, which opens the possibility of
cancellations. More generally, one can often evade those bounds by relatively
minor modifications of the model. In case of $g_\mu-2$, there's an additional
ambiguity due to the $\sim 2\sigma$ discrepancy between SM `predictions' based
on different data sets, as mentioned in the discussion of Fig.~1.

Including the DM constraint (\ref{md_1}) is fine in principle, since the
existence of DM really has been established beyond reasonable doubt in the
framework of mSUGRA or similar models.\footnote{If one entertains the idea of
  modifying gravity \cite{md_mond} rather than introducing Dark Matter,
  supergravity would certainly also have to be modified drastically.} However,
translating this into a bound on model parameters, or on physical masses, is
far from trivial. The usual procedure \cite{md_dn0,md_old,md_ddk1,md_other} is
to take eq.(\ref{md_2}) at face value. However, as discussed in the
Introduction, this means that one accepts a set of assumptions about the early
Universe that are presently untested, and indeed are difficult to test except
through Dark Matter physics. Nevertheless, since these assumptions are indeed
reasonable, it makes sense to try and see where they lead us.

It was hoped originally that the upper bound on the DM relic density (the
so--called `overclosure' constraint) would allow to establish reliable, useful
upper bounds on sparticle masses. Eq.(\ref{md_2}) shows that the relic density
goes like the inverse of the (effective) $\tilde \chi_1^0$ annihilation cross
section, which in turn (through dimensional arguments, or by unitariy
\cite{md_unit}) goes like the inverse square of the relevant mass
scale. Indeed, unitarity does allow to establish an upper bound on the mass
of any WIMP; however, this bound exceeds 100 TeV \cite{md_unit}, and is thus
not particularly useful, since we lack the means to build colliders that could
cover this kind of mass range. In the context of mSUGRA, it became clear quite
early on \cite{md_dn1} that very, even ``unnaturally'', large masses can be
compatible with the DM constraint (\ref{md_1}) even in standard cosmology.

\vspace*{5mm}
\begin{table}[h!]
\begin{center}
\begin{tabular}{|c||c|c|c|}
\hline
  (s)particle
  & \multicolumn{3}{c|}{mass bounds [GeV]} \\
 & \ \ \ \ \ \ \ \ Set I\ \ \ \ & \ \ \ \ Set II\ \ \ \  & \ \ \ \ \ Set III\
  \ \ \ \  \\
\hline
$\tilde \chi_1^0$ & 50 & 53 & [53, 61] \\
$\tilde \chi_1^\pm$ & 105 & 105 & [105, 122] \\
$\tilde \chi_3^0$ & 136 & 137 & [280, --] \\
\hline
$\tilde \tau_1$ & 99 & 99 & [630, --] \\
$h$ & 114 & 114 & [114, 122] \\
$H^\pm$ & 128 & 128 & [246, --] \\
\hline
$\tilde g$ & 374 & 383 & [383, 482] \\
$\tilde d_R$ & 444 & 444 & [774, --] \\
$\tilde t_1$ & 102 & 110 & [110, --] \\
\hline
\end{tabular}
\caption{Sparticle mass bounds in mSUGRA. These have been obtained by scanning
  over the entire allowed parameter space, defined by $m_t \in [171 \, {\rm
  GeV}, \, 185 \, {\rm GeV}], \ (m_{\tilde t_1} + m_{\tilde t_2})/2 \leq 2$
  TeV, the lower bounds on sparticle and Higgs masses from collider
  experiments, the constraint on $b \rightarrow s \gamma$ implemented as in
  \cite{md_ddk2}, simple `CCB' constraints \cite{md_ccb}, and a conservative
  interpretation of the constraint from 
  $g_\mu-2$ (essentially the overlap of the $2\sigma$ regions using $\tau$
  decay and $e^+e^-$ collider data \cite{md_ddk2}). Set II adds the DM
  constraint (\ref{md_1}), implemented using (\ref{md_2}), to the above set of
  constraints. Set III is like Set II, except that the scanned region as been
  artifically limited to the $h-$pole region, where $m_{\tilde \chi_1^0} \leq
  m_h/2$. Only {\em lower} bounds are listed for Sets I and II, while for Set
  III the allowed range is given; a dash (--) means that the upper bound is
  directly set by the arbitrary upper bound on the average stop mass.}
\label{tab:a}
\end{center}
\end{table}

Nevertheless it is clear from Figs.~1 and 2 that this constraint does exclude
very large chunks of otherwise allowed parameter space. One might therefore
think that it would at least affect the lower bounds on sparticle masses
significantly. The Table below shows that this is not really the case. This
table lists lower bounds on the masses of some new (s)particles in mSUGRA,
first without (set I) and then with (set II) including the DM constraint. In
particular, the lower bounds on many new (s)particles simply coincide with the
bounds established by collider experiments. This is true for the lighter
chargino, stau and scalar Higgs states, and essentially also holds true for
the lighter stop. The bounds on the masses of the gluino and third neutralino
are essentially the same as that in a more general MSSM, as long as gaugino
mass unification is maintained. Clearly the DM constraint still allows some
new (s)particles to be quite light! I should emphasize, however, that usually
the lower bounds in the Table cannot be saturated simultaneously. E.g. the
lower bounds on electroweak gaugino masses are saturated at $m_0 \geq 1$ TeV,
where first generations squarks and sleptons are quite heavy. Nevertheless,
the possibility of light sparticles even in this simplest of all potentially
realistic SUSY models that allow WIMP Dark Matter should be quite encouraging
to experimenters!


Set III shows these bounds (including the DM constraint) when one confines
oneself to the $h-$pole region discussed at the end of the previous
subsection. In this case there are significant upper bounds on the masses of
all gauginos.  The reason is that one needs $2 m_{\tilde \chi_1^0} \simeq m_h
\leq 120$ GeV here. Since $m_{\tilde \chi_2^0} \simeq m_{\tilde \chi_1^\pm}
\simeq M_2 \simeq 2 M_1$ and $m_{\tilde g} \simeq M_3 \simeq 6 M_1$ in mSUGRA
(due to the assumed universality of gaugino masses at the GUT scale), one
finds quite stringent upper bounds on the masses of the lighter chargino and
gluino. These bounds can be tested at the Tevatron, if it manages to collect
several fb$^{-1}$ of data; testing the gluino mass bound at the LHC is
essentially trivial. On the negative side, we see that most scalars have to be
quite heavy in this region. This is required to satisfy the $h$ search limit
from LEP with small $m_{1/2}$.

\subsection{Heavy higgsino Dark Matter}

In the last subsection I imposed the naturalness constraint that the average
stop mass should not exceed 2 TeV. If this constraint is relaxed, the deep
hyperbolical branch region becomes allowed. For $m_0 \sim 2 m_{1/2} \geq 6$
TeV (for $m_t = 178$ GeV) one finds \cite{md_ccdkr} DM--allowed points where
the LSP is an almost pure higgsino with mass near 1 TeV \cite{md_swedes}. This
region is impossible to probe even at the LHC, since squark and gluino masses
lie near 10 TeV or even higher; in fact, $m_0$ and $m_{1/2}$ could even be
exponentially larger than the weak scale! A decisive test will therefore have
to make do with the three relatively light sparticles, i.e., the three
higgsino--like states $\tilde \chi_1^0, \, \tilde \chi_1^\pm$ and $\tilde
\chi_2^0$. Currently the best bet for producing these sparticles in detectable
quantities seems to be the futuristic $e^+e^-$ collider CLIC \cite{md_clic},
which may operate at center--of--mass energy $\sqrt{s} = 3$ TeV. However,
since the mass splittings between these higgsino states amount to at most a
few GeV, detecting them will not be straightforward. The reason is that one
expects large fluxes of beamstrahlung photons at future $e^+e^-$ colliders,
which tend to get worse with increasing beam energy and increasing luminosity.
(Note that the luminosity should exceed that of LEP by a factor of order
$s_{\rm CLIC}/s_{\rm LEP} \sim 200$ to produce a useful number of $e^+e^-$
annihilation events.) This flux might be so large that each annihilation event
is accompanied by an ``underlying event'' consisting of one or more soft
$\gamma \gamma \rightarrow$ hadrons reactions \cite{md_dg}.

This can be overcome by demanding that the higgsinos are produced together
with a hard photon, that is emitted at large angle to the beam. A background
event would then have to contain a hard $e^+e^- \rightarrow \nu \bar \nu
\gamma$ annihilation plus an underlying event that happens to look like the
soft $\tilde \chi_2^0$ or $\tilde \chi_1^\pm$ decay products. A recent
calculation \cite{md_ccdkr,md_talk} indicates that this background (and
similar ones occurring at higher orders in electroweak couplings) can be
overcome, especially if longitudinally polarized beams are available.

\section{Nonminimal scenarios}

The small size of the DM--allowed parameter space in Fig.~1 is occasionally
taken as motivation to consider extensions of the mSUGRA model. Actually, a
small allowed parameter space {\it per so} should not be considered to be
problematic. New measurements, like the constraint (\ref{md_1}), are bound to
constrain the parameter space of any theory. In fact, ideally one wants to
(over--)constrain the parameter space, so that it collapses to (almost) a
point, as in QED (considered here as interaction of electrons with photons,
i.e. a 2 parameter theory). 

Slightly more worrisome is perhaps the observation that almost everywhere in
the DM--allowed mSUGRA parameter space some particular relation between
parameters is required to hold (in the pole or co--annihilation regions),
and/or one is close to the edge of the region excluded by consistency
considerations: in the $\tilde \tau$ co--annihilation region one is close to
the region where the $\tilde \tau_1$ would be LSP, leading to a forbidden
charged stable particle, whereas the focus point / hyperbolical branch region
is close to the edge of parameter space where consistent electroweak symmetry
breaking no longer works. One might therefore ask whether more `generic'
DM--allowed regions of parameter space exist in (slight) generalizations of
mSUGRA.

The answer, I believe, is No. This is illustrated by Fig.~3, which shows the
scaled DM relic density as function of one parameter, with all other
parameters held fixed. Starting point is an mSUGRA scenario with $m_t = 178$
GeV, $m_0 = 800$ GeV, $m_{1/2}=500$ GeV, $A_0 = 0$, $\tan\beta = 10$ and $\mu
> 0$. On the red curve, which is actually still an mSUGRA curve, $A_0$ has
been increased. This reduces the mass of the lighter stop eigenstate $\tilde
t_1$; eventually $\tilde t_1 - \tilde \chi_1^0$ co--annihilation
\cite{md_stop} sets in, reducing the relic density. Alas, $\Omega_{\tilde
  \chi_1^0} h^2$ quickly {\em undershoots} the desired range, i.e. one gets
the right relic density only for a very narrow range of values of $A_0$.

\begin{figure}[h!]
\centerline{\includegraphics[height=.4\textheight]{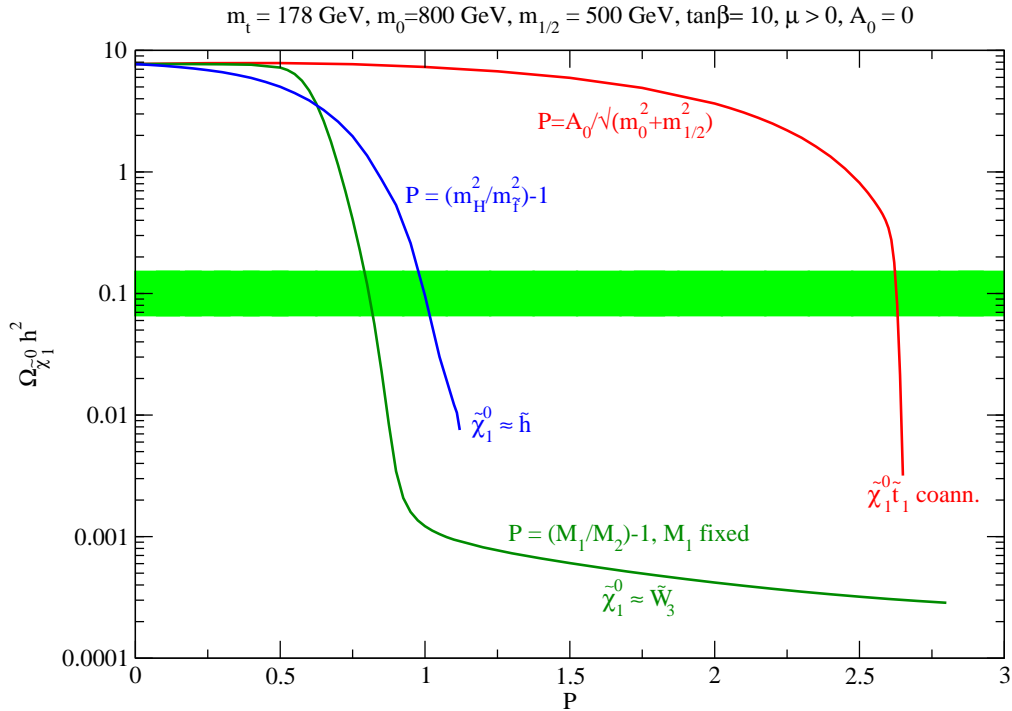}}
\caption{The scaled relic density in slight generalizations of mSUGRA. The
  desired range (\ref{md_1}) is indicated by the light green band. Demanding
  that the relic density falls in this band singles out a narrow sliver of the
  otherwise allowed range of the parameter $P$ that is being varied. See the
  text for further details.}
\end{figure}

The situation is similar on the blue line, where I have increased both Higgs
soft breaking masses at the GUT scale relative to the sfermion masses
\cite{md_nu1}. This allows to reduce the value of $|\mu|$ at the weak scale,
making the LSP more higgsino--like. However, again the relic density quickly
falls below the desired range; as we saw in the last Subsection of the
previous Section, a higgsino--like LSP would need a mass near 1 TeV to make a
good thermal CDM candidate; this is well above the LSP masses considered in
Fig.~3. The DM constraint therefore again singles out a very narrow strip of
the (extended) parameter space. Note also that in this, as well as the
previous, case the DM--allowed region is again quite close to the upper bound
on the parameter that is being varied (due to the experimental bound on
$m_{\tilde t_1}$ and on $m_{\tilde \chi_1^\pm}$, respectively).

The dark green curve is for a model where gaugino masses are not universal at
the GUT scale \cite{md_nu2}. Specifically, I have kept $M_1 = M_3$, but
reduced $M_2$, making the LSP more wino--like. Unfortunately the curve yet
again falls quickly through the desired range of relic density, i.e. the DM
constraint is again satisfied only in a narrow sliver of parameter space; in
this case the allowed region is not close to the edge, though.

On can also consider generalizing the cosmological framework. All curves in
Fig.~3 still use the standard assumptions (sufficiently high temperature, only
SM particles contribute to the Hubble expansion, no entropy production after
decoupling of the LSP). For example, a quintessence field could significantly
alter the Hubble parameter during decoupling. As far as I know, only scenarios
with increased $H$ have been analyzed so far \cite{md_quint}. In this case one
can increase the relic density by a large factor, e.g. allowing higgsino Dark
Matter with ``natural'' mass of a few hundred GeV. However, one needs to tune
the model to have a large change of $H$ during LSP decoupling and in the
present period (to account for the observed accelerated expansion of the
Universe), while keeping $H$ very close to its standard value during Big Bang
nucleosynthesis.

Another way to increase the LSP density is to assume that there were
non--thermal production mechanisms \cite{md_nontherm}. Finally, if the
lightest neutralino is long--lived but eventually decays into a very weakly
coupled superparticle, e.g. an axino \cite{md_axino} or gravitino
\cite{md_gravitino}, one automatically becomes more sensitive to details of
early universe cosmology. The simplest assumption is that the decay of the
neutralinos (or whatever else is the lightest sparticle in the visible sector)
is the by far dominant contribution to the relic density of this superweakly
interacting sparticle \cite{md_feng}. However, in general there can be
significant production of gravitinos or axinos also from the thermal bath
\cite{md_ltalk}. Finally, there might again be additional non--thermal
production mechanisms \cite{md_rouzbeh}. Scenarios with very weakly
interacting dark matter particles therefore allow to make any point in the
parameter space of mSUGRA (or similar models) DM--compatible. If the lightest
neutralino remains the lightest visible sparticle, these models often have
exactly the same collider phenomenology as if $\tilde \chi_1^0$ indeed was the
LSP.

\section{Summary and conclusions}

The constraint (\ref{md_1}) greatly constrains the parameter space of any
particle physics model of Dark Matter. The mSUGRA model still works fine,
although some relation between parameters seems to be required. One
(resurrected) possibility is $2 m_{\tilde \chi_1^0} \simeq m_h$ (the $h-$pole
region).  More generally, in mSUGRA the DM constraint does not increase lower
bounds on sparticle and Higgs masses significantly. Very (almost arbitrarily)
large gaugino and sfermion masses are allowed if the LSP is higgsino--like; it
might nevertheless be possible to test this region of parameter space at CLIC.
Generalizations of mSUGRA allow to reduce the DM relic density, but generally
do not increase the fraction of DM--allowed parameter space. Finally, if the
LSP resides in the hidden sector, pretty much any visible sector spectrum can
be made DM--safe, at the cost of predictivity and testability. I conclude that
the Dark Matter constraint is a great tool for model builders, but should
probably not be taken too seriously by collider physicists.


\subsubsection*{Acknowledgments}
I thank the organizers for inviting me to this nice conference, and the Center
for Theoretical Physics at Seoul National University for hospitality while I
was writing up this talk.

\end{document}